\documentclass[twocolumn,tighten]{aastex631}

\usepackage{amsmath,amstext,amssymb}
\usepackage{mathrsfs}
\usepackage{float}

\usepackage{afterpage}
\usepackage{xcolor}
\usepackage{threeparttable}

\usepackage{graphicx}
\usepackage{epsfig}

\usepackage{booktabs}
\usepackage{multirow}

\usepackage{accents}

\begin{document}

\title{Theoretical Foundation of Black Hole Image Reconstruction using \texttt{PRIMO}}

\author[0000-0003-1035-3240]{Dimitrios Psaltis}
\affiliation{School of Physics, Georgia Institute of Technology, 837 State St NW, Atlanta, GA 30332, USA}

\author[0000-0003-4413-1523]{Feryal \"Ozel}
\affiliation{School of Physics, Georgia Institute of Technology, 837 State St NW, Atlanta, GA 30332, USA}

\author[0000-0003-2342-6728]{Lia~Medeiros}
\altaffiliation{NASA Hubble Fellowship Program, Einstein Fellow}
\affiliation{Department of Astrophysical Sciences, Peyton Hall, Princeton University, Princeton, NJ, 08544, USA}

\author[0000-0003-3234-7247]{Tod R.\ Lauer}
\affiliation{NSF National Optical Infrared Astronomy Research Laboratory, Tucson, AZ 85726, USA}

\begin{abstract}
A new image-reconstruction algorithm, \texttt{PRIMO}, applied to the interferometric data of the M87 black hole collected with the Event Horizon Telescope (EHT), resulted in an image that reached the native resolution of the telescope array. \texttt{PRIMO} is based on learning a compact set of image building blocks obtained from a large library of high-fidelity, physics-based simulations of black hole images. It uses these building blocks to fill the sparse Fourier coverage of the data that results from the small number of telescopes in the array. In this paper, we show that this approach is readily justified. Since the angular extent of the image of the black hole and of its inner accretion flow is finite, the Fourier space domain is heavily smoothed, with a correlation scale that is at most comparable to the sizes of the data gaps in the coverage of Fourier space with the EHT. Consequently, \texttt{PRIMO} or  other machine-learning algorithms can faithfully reconstruct the images without the need to generate information that is unconstrained by the data within the resolution of the array. We also address the completeness of the eigenimages and the compactness of the resulting representation. We show that \texttt{PRIMO} provides a compact set of eigenimages that have sufficient complexity to recreate a broad set of images well beyond those in the training set.
\end{abstract}


\section{Using PRIMO to reconstruct EHT images}

Observations of nearby black holes with the EHT (Event Horizon Telescope) have generated the first high-resolution images and polarization maps at the scales of their event horizons~\citep{EHT_M87_1,EHT_M87_POL,EHT_SGRA_1}. In order to achieve such resolution, the EHT array spans the entire globe, with individual stations located in widely separated areas from Hawai'i to the French Alps and from Greenland to the South Pole. Because of the large separation between the stations, the EHT operates as a Very-Long Baseline Interferometer (VLBI), with data recorded and timetagged in each location individually and then correlated in post processing~\citep{EHT_M87_II}.

\begin{figure*}
    \centering
\includegraphics[width=0.3\linewidth]{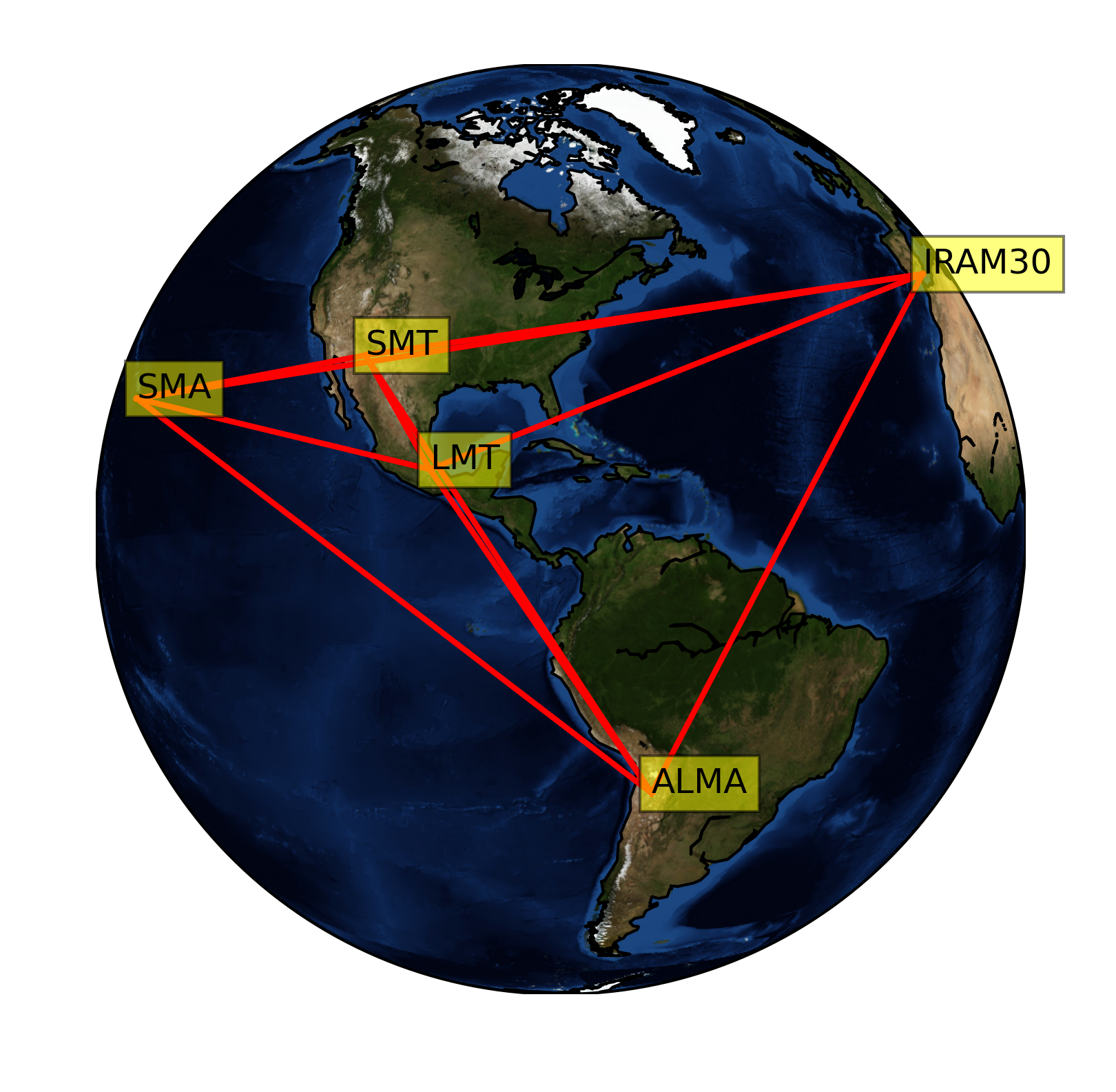}
\includegraphics[width=0.33\linewidth]{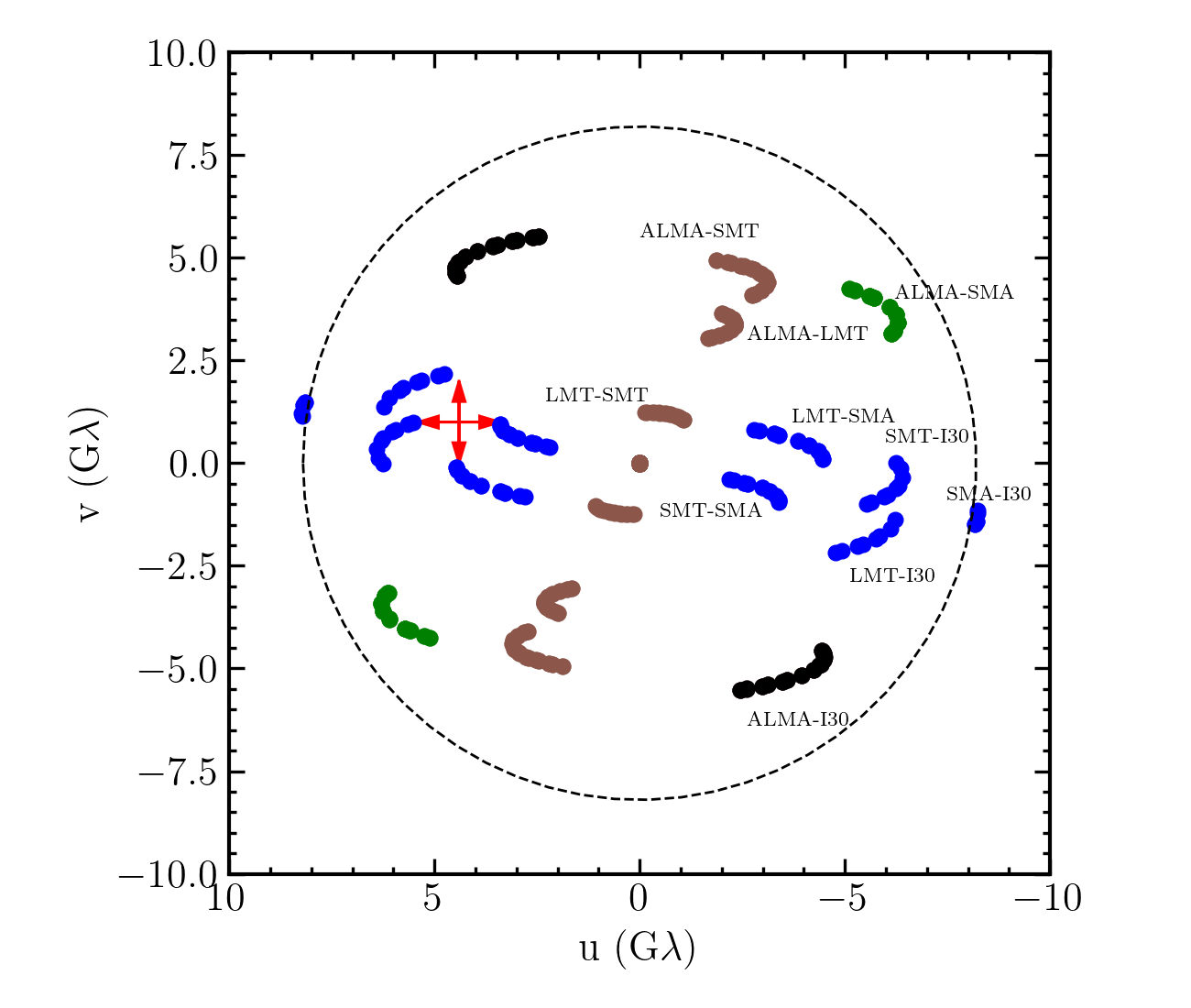}
\includegraphics[width=0.3\linewidth]{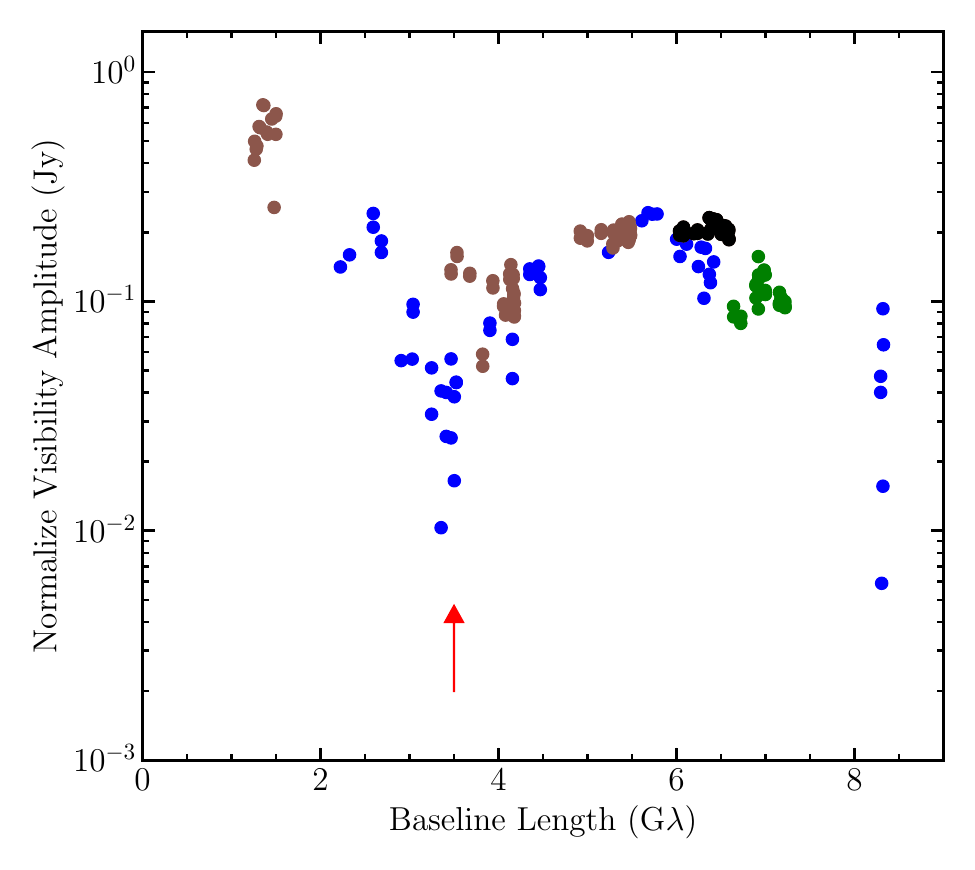}
\caption{{\em (Left)\/} The EHT locations that participated in the 2017 April 11 observations of the black-hole in the M87 galaxy. {\em (Center)\/} The coverage of the interferometric Fourier plane provided by the 2017 EHT array. The dashed circle at 8.2~G$\lambda$ encloses the region where EHT data lie. Each cluster of points corresponds to a different pair of telescope stations with individual points in each cluster representing the evolution of the baselines throughout an observing night. The clusters of points are separated by at least $\sim 2$~G$\lambda$, as indicated by the sizes of the red arrows. The large geographic separation between the EHT stations gives rise to a very sparse coverage of the Fourier plane. {\em (Right)\/} The observed visibility amplitudes as a function of baseline length. The location of the right arrow shows the sharp deep minimum at a baseline length $\sim 3.5$~G$\lambda$ \label{fig:eht}}
\end{figure*}

The correlated EHT data provide the complex Fourier components of the observed image at a distinct set of spatial frequencies set by the lengths and orientations of the piecewise separations between its telescope stations. These Fourier components need to be inverted in order to generate the observed images. However, because of the large separation between the stations, an EHT dataset provides only very sparse Fourier-plane coverage, necessitating the use of algorithms to fill in the gaps when reconstructing the images.  

The initial images of the black holes in the center of the M87 and of the Milky Way galaxies were reconstructed using primarily regularized maximum likelihood methods as well as more traditional approaches based on the \texttt{CLEAN} algorithm~\citep{EHT_M87_IV,EHT_SGRA_III}. In both cases, they revealed the presence of a bright ring of emission surrounding the shadow of the black hole, constrained our understanding of plasma physics at horizon scales, and led to new tests on the predictions of Einstein's general theory of relativity~\citep{Psaltis2020,EHT_SGRA_VI}.

We recently developed \texttt{PRIMO}~\citep[Principal-component Interferometric Modeling;][]{Medeiros2023a} for interferometric image reconstruction and used it to obtain a high-fidelity image of the M87 black hole from the 2017 EHT data~\citep{Medeiros2023b}. In this approach, we decompose the image into a set of eigenimages, which the algorithm ``learned'' using a very large suite of black-hole images obtained from general relativistic magnetohydrodynamic (GRMHD) simulations. The algorithm then obtains the optimal linear superposition of these eigenimages by minimizing a loss function between the Fourier components of the model image and the observations. The high fidelity of the \texttt{PRIMO} reconstruction is underscored by the fact that the Fourier components of the reconstructed image are statistically consistent with the observed data without requiring {\em ad hoc} regularizers~\citep{Medeiros2023b}.

Relying on the training set to reconstruct the observed image, however, naturally leads to two important questions: {\em (i)\/} Is the determination of a unique image exposing a lack of diversity in the particular training set that was used, i.e., does the algorithm select among a narrow range of possibilities? {\em (ii)\/} Are the assumptions inherent in the simulations that generated the training images imprinted on reconstructed images? In summary, was the algorithm biased to find in the image the features it already ``knew'' were there? In this paper, we aim to address these two questions.

\section{EHT Observations Are Not As Sparse As You Think}

In the sections that follow, we will use a variety of analytic and numerical arguments to address the reservations outlined above. However, we start with a more qualitative approach, first to introduce the crucial concept that interferometric observations in general may be less sparse, less incomplete, than a naive examination of their standard visibility maps would suggest.  This in turn means that \texttt{PRIMO} and other similar image reconstruction techniques can more easily recover the intrinsic structure of the accreting black hole than might first appear to be the case.

The center panel of Figure \ref{fig:eht} shows a classic map of the interferometric $u-v$ plane of the EHT observations of M87 taken over an example night of observations. Each interval of observations for each baseline is plotted as a point.  While, in a broad sense, the coverage over the $u-v$ plane is fairly symmetric over a good range of angular scales, most of the plane is empty.  Inferring the form of the missing object structure in these empty areas is the challenge faced by all image reconstruction techniques. The small fraction of the map occupied by the actual observations might suggest that there is considerable freedom in how the missing data may be filled in, thus raising serious concerns about the uniqueness of any reconstructed image.

In practice, however, the compact angular extent of the M87 black hole and of its inner accretion flow, which is no more than an order of magnitude larger than the limiting angular resolution of the EHT, means that the Fourier components at any location in the map will be highly correlated with neighboring components. The length of the correlation scale in the $u-v$ plane is, in fact, set entirely by the overall image angular scale. For both EHT targets, this is large enough to partially fill in much of the gaps in the Fourier coverage, reducing substantially the effective number of degrees of freedom in the model image down to the resolution of the EHT array. 

\begin{figure}[t]
    \centering    \includegraphics[width=0.99\linewidth]{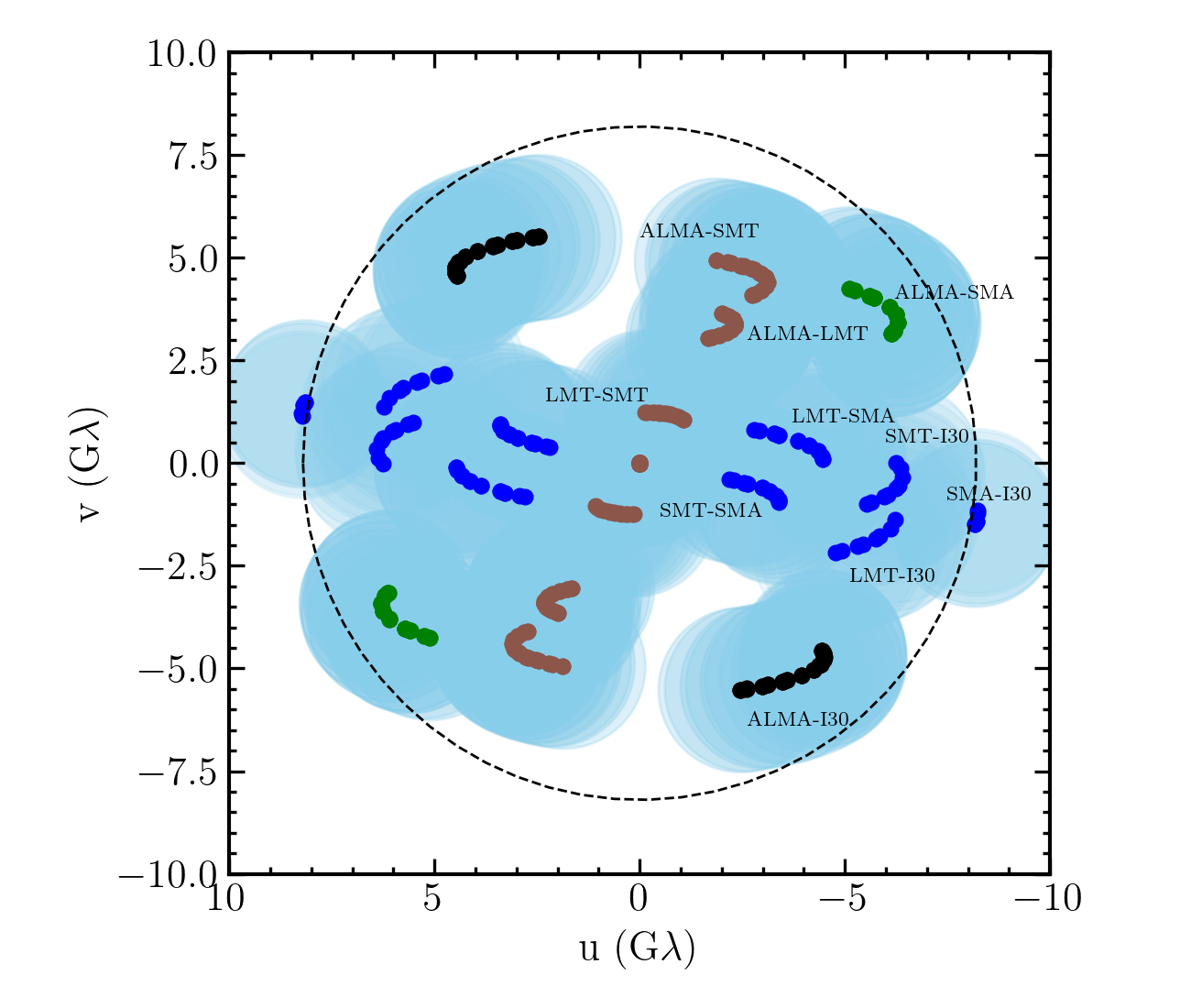}
    \caption{The UV-plane coverage of the 2017 April 11 EHT observations of the M87 black-hole shown in Figure \ref{fig:eht}, but now with cyan coloring to identify areas of the plane that are highly correlated with the observations. The origin and form of the correlations will be discussed  in detail in $\S\ref{sec:cor}.$ Cyan circles indicating the 2G$\lambda$ radii correlation scale are centered on each visibility point.  The dashed circle at 8~G$\lambda$ outlines the region of $u-v$ space inside which \texttt{PRIMO} can exploit these correlations to provide good azimuthal coverage of the Fourier plane.}
    \label{fig:uv}
\end{figure}

Figure~\ref{fig:uv} qualitatively shows how the visibility correlation scale reduces considerably the truly unobserved regions of the Fourier domain interior to the EHT angular band-limit. This is why a small number of \texttt{PRIMO} eigenimages are sufficient to reconstruct the black-hole images. We will explore the form and origin of the visibility correlation in detail in $\S\ref{sec:cor},$ but it readily emerges in all cases where the extent of the object in space is limited, or strongly ``windowed," to use the more standard technical term.  Consider an infinite pure sine-wave multiplied by finite window, or ``structure function'' as we will later call it, that resembles the overall footprint of the accreting black hole.  Multiplication in space of an extended object by a compact window is equivalent to the {\it convolution} in Fourier space of the object and window transforms.  As the pure sine-wave transform is a pair of delta-functions, it's instantly evident that the Fourier transform of the windowed sine-wave would then comprise two window transforms centered on the locations of the original delta functions. The sharp visibility points in Figure \ref{fig:eht} are more correctly represented as overlapping oblate objects, as shown in Figure \ref{fig:uv}.

Given all of these considerations, it is clear that the unknown or empty regions of the $u-v$ plane seen in Figure~\ref{fig:eht} are actually heavily constrained over much of the domain, depending on the details of the correlation kernel and the distance to the nominal centers of the visibility observations.  Unobserved areas in the $u-v$ plane far removed from the locations of the observations do have some freedom in allowable amplitudes --- but not arbitrarily so.  This in turn means that the repair-work done by algorithms like \texttt{PRIMO} is considerably more modest than the $u-v$ plane representation in Figure~\ref{fig:eht} would suggest.

\section{Imaging Black Holes with the Event Horizon Telescope}

In this section, we provide a quantitative analysis of the correlation lengths in the Fourier plane of interferometric observations. As an interferometer, the EHT measures the correlated flux between pairs of telescope stations separated by large distances. According to the van Cittert-Zernike theorem, this correlated flux for a pair of stations is equal to the complex Fourier component of the image brightness
\begin{equation}
{\cal V}(u,v)\equiv\int\int e^{-2\pi(x u +y v)}I(x,y)\; dx dy\;,
\label{eq:vCZtheorem}
\end{equation}
evaluated at a 2D-spatial frequency $(u,v)\equiv\vec{b}/\lambda$. Here the brightness distribution of an image on the sky is $I(x,y)$, where $x$ and $y$ are angular coordinates; $\vec{b}$ is the displacement vector between the two stations in the array, projected orthogonally to the line of sight to the source, and $\lambda=1.3$~mm is the wavelength of the observation. Using standard terminology, we will also refer to ${\cal V}(u,v)$ as the complex visibility and to the spatial frequencies $(u,v)$ as the baselines of the array, with the latter measured in $G\lambda$. Note that, because the image brightness is a real function, $\vert V(-u,-v)\vert=\vert V(u,v)\vert$.

Figure~\ref{fig:eht} shows the locations on the globe of the 5 geographical locations that participated in the 2017 EHT observations of the M87 black hole, as viewed from the direction of the source~\cite[see][for the details of the array and the telescopes involved]{EHT_M87_II}. The baselines connecting pairs of stations correspond to the frequencies in the Fourier $(u,v)$ plane at which EHT measures the complex visibilities. The largest baseline of this configuration is $\sim 8.2~$G$\lambda$, corresponding to the separation between the IRAM~30~m telescope in Spain and the SMA/JCMT telescopes in Hawai'i. 

Each pair of stations provides a single baseline at a given time, which then evolves during the course of the night with the rotation of the Earth. This is the reason why each pair of stations generates a cluster of points that lie on the arc of an elliptical curve in the $(u,v)$ plane. For the 2017 M87 observations, these arcs have rather small extents. Importantly, they are separated by gaps that are devoid of data. The sizes of these gaps are $\gtrsim 2$~G$\lambda$. It is this sparsity of the coverage of the Fourier $(u,v)$ plane that introduces the biggest challenge in image reconstruction with EHT data.

In order to overcome this challenge, \texttt{PRIMO} decomposes the image into a linear combination of a small number ($N_i\simeq 20$) of eigenimages~\citep{Medeiros2023a,Medeiros2023b}. 
The coefficients of these sums are obtained using an optimization algorithm such that the model Fourier maps are in agreement with the data based on an appropriately defined loss function.

In principle, the set of eigenimages could be any basis set of two-dimensional functions, such as the Zernike polynomials. Even though this would ensure completeness, it would not necessarily provide a compact representation of the images. Indeed, images with sharp edges, such as those produced at the boundary of a black-hole shadow~\citep{Psaltis2015}, would require a very large number of smooth Zernike polynomials to be reproduced. (This is a generalization of the Gibbs phenomenon in spectral decomposition of functions). Instead, in \texttt{PRIMO}, we opted for the algorithm to learn a compact set of basis functions (or eigenimages), by training it on a very large number of realistic simulated black-hole images. In other words, we aimed for a sparse-representation modeling of the images using a linear combination of atoms from a dictionary, with the atoms obtained from a well chosen training set~\citep{Rubinstein2010}. 

We obtained the dictionary atoms via a linear Principal Component Analysis (PCA) of the training set. We could have chosen instead a non-linear PCA algorithm, such as an autoencoder~\citep{Kramer1991}, to perform this task. However, the linearity of the traditional PCA decomposition ensures that the atoms of the basis set in the Fourier domain, where we compare the model images to the data, are just the Fourier transforms of the eigenimages of the basis set in the image domain, where we calculate the simulated images~\citep{Medeiros2018}. The orthogonality of the PCA basis reduces possible correlations between the coefficients of the decomposition since no eigenimage can be written as a linear combination of the other eigenimages. 

The compactness of the dictionary, i.e., the number $N_{\rm i}$ of PCA components necessary to reproduce the EHT images, is determined by the error budget of the EHT data. For the 2017 dataset, requiring that the residual of projecting the training set of images onto $N_{\rm i}$ PCA components is at most comparable to the $\sim 1$\% error in the data allowed us to set $N_{\rm i}\simeq 20$~\citep{Medeiros2023a}. 

In the following sections, we explore the reasons why the combination of the PCA decomposition of the training set of simulated black-hole images with the particular properties of the coverage of the interferometric $u-v$ plane for the 2017 EHT observations of M87 allow for a robust reconstruction of this black-hole image.

\begin{figure*}[t]
    \centering
\includegraphics[width=0.9\linewidth]{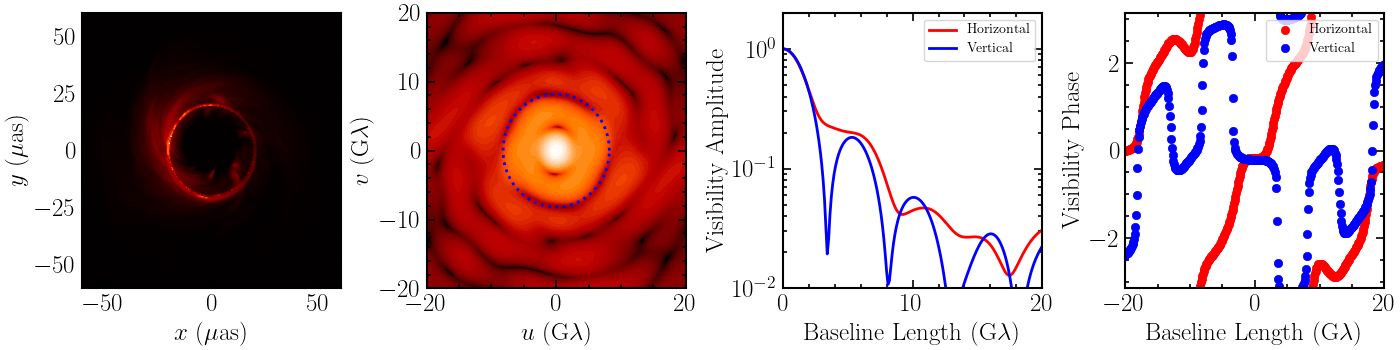}
\includegraphics[width=0.9\linewidth]{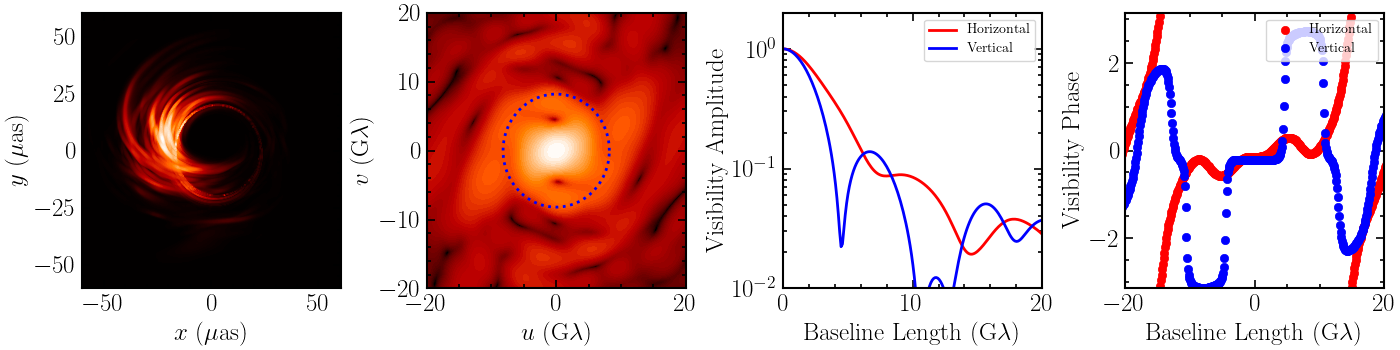}    \includegraphics[width=0.9\linewidth]{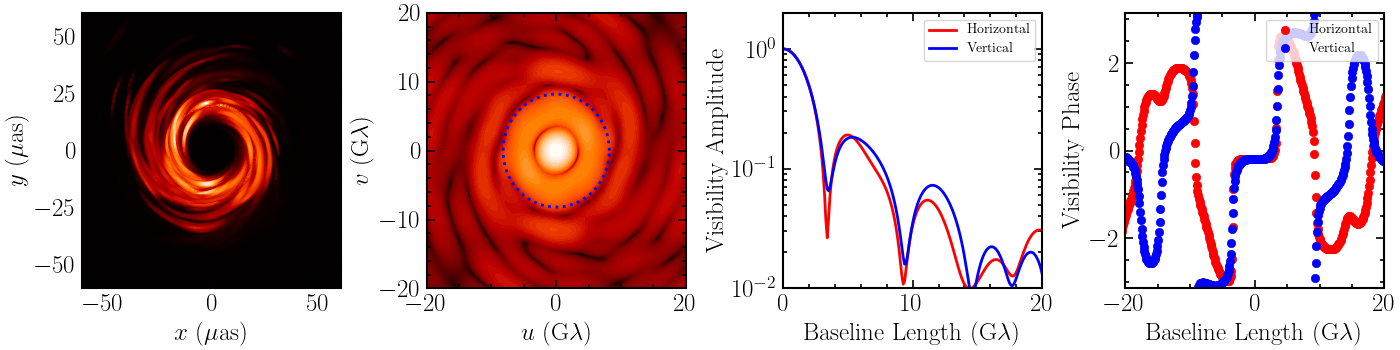}
    \caption{Three examples of simulated black-hole images and their Fourier maps for parameters that resemble the 2017 EHT data of M87. The left panels show the simulated images, the second panels show the interferometric visibility amplitudes, and the two rightmost panels show vertical and horizontal cross sections of the visibility amplitudes and phases of each image. The dashed circles in the second panels represent the largest baseline length in the 2017 EHT observations of M87 at 8.2~G$\lambda$. The complex visibilities of all images show significant correlations in the $u-v$ plane with a characteristic scale that is comparable to the baseline length that resolves each image (i.e., about 2-3~G$\lambda$ in the case of the simulated images).}
    \label{fig:examples}
\end{figure*}

\section{Interferometric Fourier Maps and Resolving Baselines for Compact Images}

The ability of an image reconstruction algorithm to fill in the Fourier components of the underlying image in the gaps between the $(u,v)$ locations of the measurements determines its performance. In principle, the amplitudes and phases of the Fourier components are arbitrary and unconstrained in the gaps where there are no data. In practice, however, this is not the case, as the Fourier components of nearby $(u,v)$ locations are highly correlated for any compact image. In this section, we first use a few examples of simulated black-hole images in order to illustrate the correlations in the Fourier maps of the images. In the following section, we then use numerical examples and analytic arguments to understand these correlations.

Figure~\ref{fig:examples} depicts three sample simulated black-hole images, their 2-dimensional Fourier transforms, and two cross sections of the latter, broken into amplitude and phase. These images were chosen randomly from the large simulation suite reported in~\citet{EHT_M87_V}, with image sizes and fluxes similar to those expected for the M87 black hole. All these images are compact and have finite sizes, i.e., the brightness of each image drops practically to zero outside some inner region. This results in visibility amplitudes that drop with increasing baseline length. Hereafter, we will denote by $u_0$ and $v_0$ the baseline lengths at which a horizontal and a vertical cross section of the visibility amplitude drops to half its central value. In standard terminology, these are the baseline lengths that ``resolve'' the image along these two orientations. By construction, these baseline lengths corresponds to the angular size $\theta_{\rm im}$ of the image along each orientation, i.e., $u_0\sim \theta_{\rm im}^{-1}$ in the horizontal orientation. For the images shown in Figure~\ref{fig:examples}, these baseline lengths are $\simeq 2-3$~G$\lambda$.

Although each image shows the characteristic black-hole shadow, the morphology of the bright emission that surrounds it is widely different between the three images. This large difference is also reflected in the corresponding Fourier maps, especially at baselines $\gtrsim 10$~G$\lambda$, which are beyond the reach of the EHT and are affected by the finer image features. However, even at these large baselines, the characteristic scale over which the visibilities vary appears to be comparable to the size of the central peak, i.e., of order the baseline lengths that resolve the images. This implies that the complex visibilities in nearby locations in Fourier space are highly correlated. Visually, the widths of the red ripples in the Fourier maps (central panels in Fig.~\ref{fig:examples}) are comparable to the sizes of the central bright blobs. Similarly, the widths of the peaks in the cross sections (right panels in the Figure) are similar to that of the central peak.

Are the three examples shown in Figure~\ref{fig:examples} representative of a more general phenomenon or are they specific to our training set? We address this question quantitatively in the following section.

\section{Fourier-Space Correlations of Interferometric Data for Compact Images}\label{sec:cor}

In this section, we demonstrate that the correlation length in the Fourier map of a compact image (or, equivalently, the characteristic baseline scale over which the Fourier components vary) is comparable to the baseline scale that resolves the image itself. 

We first explore analytically the extreme case in which a compact image of finite size $\theta_{\rm im}$ has fine structure that is described by white noise, i.e., the fine structure has zero correlation length. To achieve this, we decompose the image brightness on the sky, $I(x,y)$, into a product
\begin{equation}
    I(x,y) = I_{\rm fs}(x,y) \cdot W(x,y)
    \label{eq:decomp}
\end{equation}
of a 2-dimensional function $I_{\rm fs}(x,y)$ that extends to infinity and describes the underlying fine structure in the image (the white noise) and an object shape function $W(x,y)$. The latter tapers quickly at angular distances larger than the angular size of the image, i.e., for $x,y\gtrsim \theta_{\rm im}$. As discussed earlier, the characteristic scale of the Fourier map of the shape function, $\tilde{W}(u,v)$, is approximately equal to the baseline $u_0$ that resolves the image. Hereafter, tilde denotes the Fourier transform of a function and $(u,v)$ are the Fourier frequencies, or baseline lengths, in units of the wavelength of observation, as before. 

The correlation length of the Fourier map of the image can be obtained from the 2-dimensional autocorrelation function {\em in Fourier space}, i.e.,
\begin{equation}
{\cal A}(\delta u, \delta v)=\int du \int dv \tilde{I}(u,v) \tilde{I}^*(u+\delta u,v+\delta v)\;.
\label{eq:autocor}
\end{equation}
The correlation lengths along the two primary orientations are then the scales over which the autocorrelation function drops to negligible values. According to the convolution theorem, the autocorrelation function is the (inverse) Fourier transform of the square of the image, which we write symbolically as 
\begin{equation}
    {\cal A}(\delta u, \delta v)\rightleftharpoons I^2(x,y)
    \label{eq:autocorFFT}
\end{equation}
or ${\cal A}=\tilde{I^2}$. Inserting decomposition~(\ref{eq:decomp}) into equation~(\ref{eq:autocorFFT}), we obtain
\begin{eqnarray}
    {\cal A}&\rightleftharpoons& \left(I_{\rm fs} \cdot W\right)^2 \nonumber\\
    \Rightarrow {\cal A} &\rightleftharpoons& \left(I_{\rm sf}^2\right) \cdot \left(W^2\right)\nonumber\\
    \Rightarrow {\cal A} &=& \tilde{\left(I_{\rm fs}^2\right)} \circledast \tilde{\left(W^2\right)}\nonumber\\
    \Rightarrow {\cal A} &=& {\cal A}_{\rm fs} \circledast \tilde{\left(W^2\right)}\nonumber\\
    \Rightarrow {\cal A} &=& \tilde{\left(W^2\right)}\;.
    \label{eq:autocor_conv}
\end{eqnarray}
In the last two steps of the derivation we used the fact that, by assumption, the Fourier transform of the square of the image fine-structure is equal to the autocorrelation function of white noise and that the latter is just a delta function. The last expression shows that, even in the extreme case where the fine structure in the image is described by white noise, the autocorrelation function of the image in Fourier space is equal to the Fourier transform of the square of the image shape function. For an image of size $\theta_{\rm im}$, the characteristic scale of the square of its shape function is $\theta_{\rm im}/\sqrt{2}$, the characteristic scale of the Fourier transform of the latter is $\sim \sqrt{2}/\theta_{\rm im}\sim \sqrt{2} u_0$ and, therefore, the characteristic scale of the autocorrelation function is also $\sim \sqrt{2} u_0$. In other words, even though the correlation length of the Fourier transform of the fine-structure of the image is zero, the correlation length of the image itself is finite and comparable to the baseline length that resolves the image.

In order to explore the correlation lengths in the Fourier maps of realistic black-hole images, we now introduce another quantity that conveys similar information to the length scale of the autocorrelation function but is local in nature. Because it does not require the calculation of the autocorrelation function itself, this quantity is computationally easier to evaluate. Moreover, as it will become apparent at the end of this section, this quantity allows us also to devise regularizers that are not {\em ad hoc} but are, instead, motivated by the properties of the images and can be used in regularized maximum likelihood reconstruction methods. (Note that we do not need regularizers for the image reconstruction with \texttt{PRIMO}.)

Our goal is to quantify the characteristic baseline scales $u_{\rm s}$ and $v_{\rm s}$ over which the complex visibilities change in the two orientations of the Fourier domain corresponding to each image. For example, if the visibility around a point with coordinates $(u_i,v_i)$ in the Fourier domain can be expressed along the horizontal orientation as the power-law function
\begin{equation}
V(u,v) \sim V_i\left(\frac{u-u_i}{u_{\rm s}}\right)^n\;,
\end{equation}
then the characteristic scale of variation along the $u-$orientation will be
\begin{equation}
u_s \equiv \left[\frac{1}{n!}\left(\frac{1}{V_i}\frac{\partial^n V}{\partial u^n}\right)\right]^{-1/n}\;.
\end{equation}

The Fourier maps of black-hole images are expected to have a series of successive extrema (see, e.g., Fig.~\ref{fig:examples}), with the first and second derivatives of the visibilities vanishing at different Fourier frequencies. To account for this, we use the harmonic average of the scales that correspond to these first two derivatives of the visibility function such that the shortest of these two scales dominates the result. For example, for the horizontal orientation, we write
\begin{equation}
    u_{\rm s}=2\left[
    \left(\frac{1}{\langle V\rangle}
    \frac{\partial V}{\partial u}\right)+\left(\frac{1}{\langle V\rangle}
    \frac{\partial^2 V}{\partial u^2}\right)^{1/2}
    \right]^{-1}\;.
    \label{eq:scale}
\end{equation}
We use a similar expression for the evaluation of the characteristic baseline scale $v_s$ in the vertical orientation.

To calculate these characteristic baseline scales, we decompose again a compact image according to equation~(\ref{eq:decomp}). However, this time we do not assume that the fine structure can be described as white noise. By definition, the characteristic scale of the Fourier map of the shape function of the object is approximately equal to the baseline that resolves the image: 
\begin{equation}
\partial \tilde{W}(u,v)/\partial u \sim \partial \tilde{W}(u,v)/\partial v \sim \tilde{W}/u_0\;.
\end{equation}
The Fourier map of the compact image is then the convolution of the Fourier maps of the two functions
\begin{equation}
 V(u,v) = \tilde{I}(u,v) = \tilde{I}_{\rm fs}(u,v) \circledast \tilde{W}(u,v)\;.
\end{equation}
Evaluating the characteristic scale of variation in the Fourier plane using equation~(\ref{eq:scale}) and assuming that the two derivatives generate comparable contributions to the sum, we obtain
\begin{eqnarray}
    u_{\rm s}&\simeq &
    \left(\frac{1}{\langle V\rangle}
    \frac{\partial V}{\partial u}\right)^{-1} = 
    \left[\frac{1}{\tilde{I}_{\rm fs}\circledast \tilde{W}} \frac{\partial}{\partial u}
    (\tilde{I}_{\rm fs}* \tilde{W})\right]^{-1}\nonumber\\
    &=&\left[\frac{1}{\tilde{I}_{\rm fs}\circledast \tilde{W}}  (\tilde{I}_{\rm fs}\circledast \frac{\partial}{\partial u}
   \tilde{W})\right]^{-1}\nonumber\\
    &\simeq &\left[\frac{1}{\tilde{I}_{\rm fs}\circledast \tilde{W}}  (\tilde{I}_{\rm fs}\circledast \frac{\tilde{W}}{u_0})\right]^{-1}
    = u_0\;.
\end{eqnarray}
A similar expression can be derived for the characteristic scale $v_{\rm s}$ along the vertical orientation. This demonstrates that {\em the characteristic scale over which the visibilities of a compact image vary on the Fourier plane is comparable to the baseline length that resolves the image, practically independent of the structure and details of the image itself.}

\begin{figure}[t]
    \centering
\includegraphics[width=0.99\linewidth]{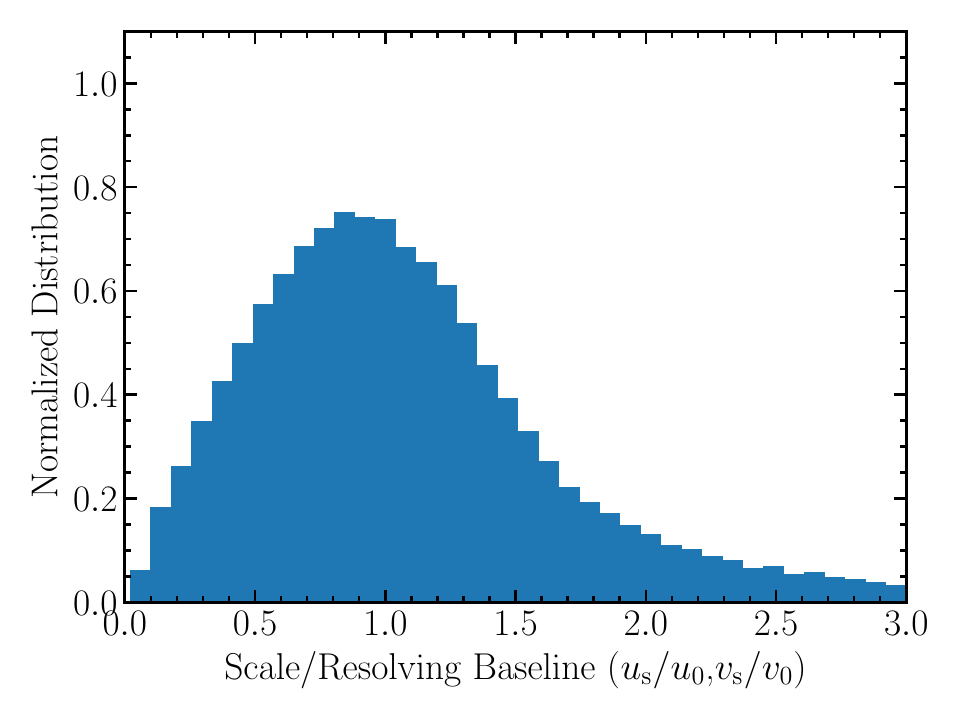}
    \caption{The combined distribution of the ratios $u_{\rm s}/u_0$ and $v_{\rm s}/v_0$ between the baseline length scales over which the complex visibilities of 100 simulated black-hole images change along the horizontal and vertical orientations and the corresponding baseline length that resolves each image. The fact that this distribution is narrowly peaked around unity implies that regions in $u-v$ space that are within one resolving baseline length of each other are highly correlated.}
    \label{fig:scales}
\end{figure}

As a final, numerical demonstration, we quantify the scale in Fourier space over which visibilities vary in realistic black-hole images obtained via GRMHD simulations. For this, we chose 100 random snapshots from simulations with different magnetic field topologies as well as different black-hole spins ($a=-0.94, -0.5, 0., 0.5, 0.94$); see \citet{EHT_M87_V} for the details of the simulations. We then used equation~(\ref{eq:scale}), as well as the equivalent equation for the vertical orientation, to calculate the characteristic length scales $u_{\rm s}$ and $v_{\rm s}$ for each image and for all regions of the Fourier plane with baseline lengths $\le 20$~G$\lambda$. Finally, we evaluated the numerical derivatives using the second-order differencing scheme:
\begin{eqnarray}
   \frac{\partial V}{\partial u}&=&
    \frac{1}{2h}\left\vert V(u+h,v)-V(u-h,v)\right\vert \nonumber\\
   \frac{\partial^2 V}{\partial u^2}&=&
    \frac{1}{h^2}\left\vert V(u+h,v)-2V(u,h)+V(u-h,v)\right\vert\nonumber\\
    \langle V\rangle &=&\frac{1}{3}\left\vert V(u-h,v)+V(u,v)+V(u+h,v)\right\vert \;,
    \label{eq:numericaldif}
\end{eqnarray}
where $h$ is the grid spacing in the Fourier domain. 

Figure~\ref{fig:scales} shows the combined distribution of the ratios of the characteristic scales along each orientation to the corresponding resolving baseline, i.e., $u_{\rm s}/u_0$ and $v_{\rm s}/v_0$. As expected from the inspection of Figure~\ref{fig:examples} and from the earlier analytic arguments, this combined distribution is narrowly concentrated at a value close to unity, demonstrating again that the characteristic scale of variation of the visibilities in the Fourier plane is set by the baseline length that resolves each image. As a result, generating a dictionary of eigenimages from a training set of compact images with appropriately chosen Fourier-space correlation lengths enables an algorithm like \texttt{PRIMO} to take advantage of these correlations and generate an image from sparse data.

Image reconstruction with \texttt{PRIMO} does not require regularizers in order to converge to a physical solution. The results discussed in this section provide us, nevertheless, with a natural choice of regularizer that one could use in maximum likelihood methods. Indeed, instead of penalizing large gradients of brightness on the image plane~\citep{Candes2005}, which is not a good requirement for images of black-hole shadows, a maximum likelihood method can penalize correlation lengths in Fourier space that are significantly larger than the resolving baseline length for each image. In practice, for each candidate image reconstruction one would obtain the baseline lengths $u_0$ and $v_0$ along two orientations at which the visibility amplitude drops to half its central value. Then they could apply equation~(\ref{eq:scale}) with the differencing scheme~(\ref{eq:numericaldif}) to calculate the characteristic baseline scales $u_s$ and $v_s$ and use them to form the regularizer \cite[cf.][]{Chael2016}
\begin{equation}
    S=\alpha \left[\left(\frac{u_s}{u_0}\right)^2+\left(\frac{v_s}{v_0}\right)^2\right]\;.
\end{equation}
Here, the parameter $\alpha$ measures the strength of the regularizer. As demonstrated in this section, compact images will always be characterized by $S\lesssim 1$, independent of the details of the images themselves.

\section{Filling the interferometric gaps with \texttt{PRIMO}}

It is well understood that the resolving power of an interferometric array is determined by its largest baseline, which sets the maximum spatial Fourier frequency that can be measured. However, the results discussed in the previous section argue that the ability of an image reconstruction algorithm to fill in the gaps in the Fourier plane depends critically on a different design property of the interferometer: the ratio between the baseline that resolves the image and the typical $u-v$ separation between the various baselines in the array. 

Let us consider first the case in which the separation between nearby baselines (or more, precisely, between nearby clusters of baselines) is much smaller than the baseline that resolves the image. This implies that, for the reasons shown in the previous section, the separation between nearby baselines is much smaller than the characteristic length in Fourier space over which the visibilities change and, therefore, that the visibilities do not change appreciably between the measurements. In this case, an algorithm can simply assume a piecewise constant visibility function between the measurements and perform an inverse Fourier transform to the image plane. The data constrain fully the reconstructed image, with little input necessary from the algorithm.

In the opposite extreme, i.e., when the separation between nearby baselines is much larger than the resolving baseline, there is very little information in the measurements that can be used to fill in the gaps in the Fourier plane. At the mid points between the various baselines, the visibility function will take values that are uncorrelated with those of the measurements and, therefore, unpredictable. In this case, an image reconstruction algorithm will need to make strong {\em a priori\/} assumptions regarding the image, in order to fill in the gaps and generate an image.

In the case of the 2017 EHT data on the M87 black hole, the baseline that resolves the image in the N-S orientation is $u_0\simeq 2$~G$\lambda$ (see Fig.~\ref{fig:eht}). Unfortunately, this is the only orientation with data points at these baseline lengths. However, the high degree of azimuthal similarity between the measurements at different baselines strongly suggests that the resolving baseline is similar to $\simeq 2$~G$\lambda$ along other orientations, as well. Therefore, the characteristic baseline scale over which the visibilities are correlated is also expected to be similar to 2~G$\lambda$ for all orientations.

\begin{figure}[t]
    \centering
\includegraphics[width=0.99\linewidth]{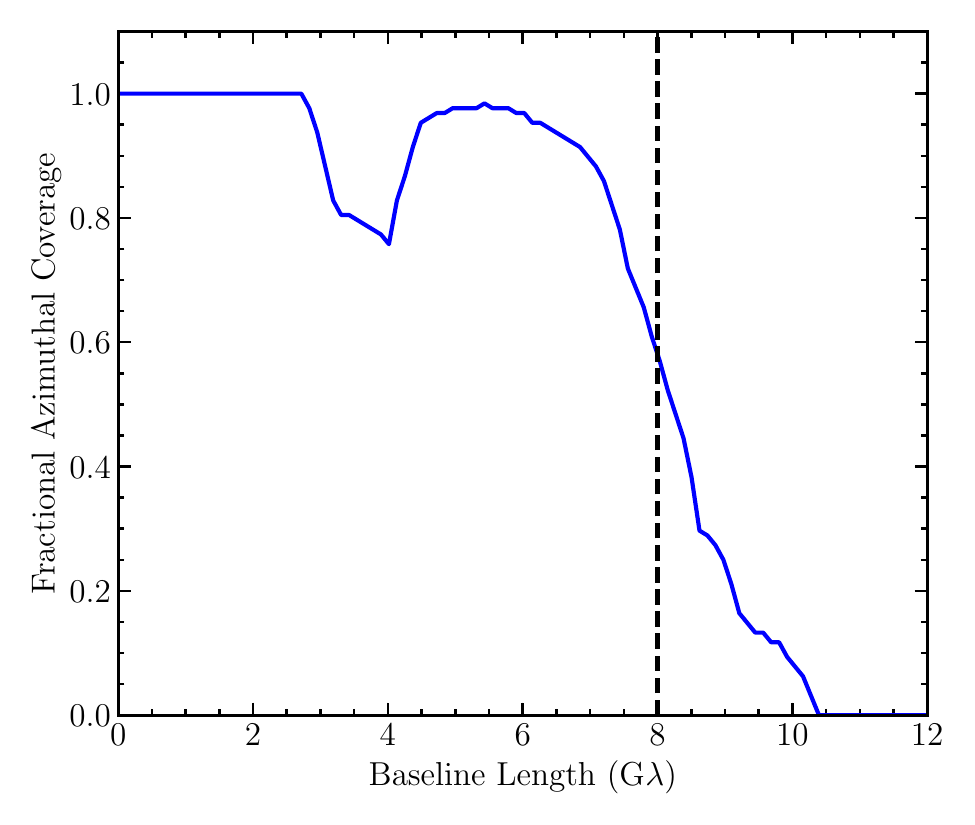}
    \caption{The fractional azimuthal coverage of regions in the interferometric $u-v$ plane for which the Fourier components of the image are highly correlated to those of the observational points, as a function of baseline length. Up to a baseline of $\sim 8$~G$\lambda$ (dashed line), \texttt{PRIMO} can exploit these correlations to provide a good azimuthal coverage of Fourier space.}
    \label{fig:uv_frational}
\end{figure}

Figure~\ref{fig:uv} uses this information to illustrate the regions in the Fourier plane that are highly correlated with the locations of the 2017 EHT measurements. The cyan filled circles in this figure are centered at the location of each baseline for which there has been a measurement during the 2017 April 11 observations. The radius of each circle is equal to the 2~G$\lambda$ characteristic baseline scale over which the visibilities are correlated. Clearly, at baseline lengths $\lesssim 8$~G$\lambda$, the typical separation between baselines, both in the radial and azimuthal orientations, is at most comparable to the correlation length, with very few exceptions along the N-S orientation. 

Figure~\ref{fig:uv_frational} further quantifies this by showing the fractional azimuthal coverage of regions in the Fourier plane that are within one correlation length of a data point, as a function of baseline length. In order to evaluate this quantity for a given baseline length $b$, we draw a circle of radius $b$ and calculate the arc length of points along its circumference for which the distance to the nearest measurement point is less than 2~G$\lambda$. Up to a baseline length of $\sim 8$~G$\lambda$, this azimuthal coverage is practically perfect suggesting that the visibilities of the M87 black-hole image can be accurately reconstructed everywhere in the Fourier plane (up to this baseline length) using the 2017 EHT observations. 

\section{Learning a nearly complete basis set of eigenimages}

In the previous sections, we have established the fact that the expected degree of correlations in the Fourier plane of the 2017 EHT observations of the M87 black hole allow, in principle, for an algorithmic reconstruction of the underlying image. The performance of any particular algorithm depends then on its ability to exploit these correlations with the high signal-to-noise data provided by the EHT. 

As shown in Figures~\ref{fig:uv} and \ref{fig:uv_frational}, the EHT data support a reconstruction of the Fourier plane up to baseline lengths of $\simeq 8$~G$\lambda$. At the same time, the $\sim 2$~G$\lambda$ correlation length between nearby baselines suggests that there are at least $(\pi 8^2)/(\pi 2^2)=16$ independent ``resolution'' elements on the Fourier plane. This is related to the reason underlying the result of \citet{Medeiros2023a} that the Fourier-plane residuals of reconstructing tens of thousands of simulated black-hole images with $N$ PCA components become negligible ($\lesssim 2$\%) when $N\gtrsim 20$.

\begin{figure*}[t]
    \centering
\includegraphics[width=0.45\linewidth]{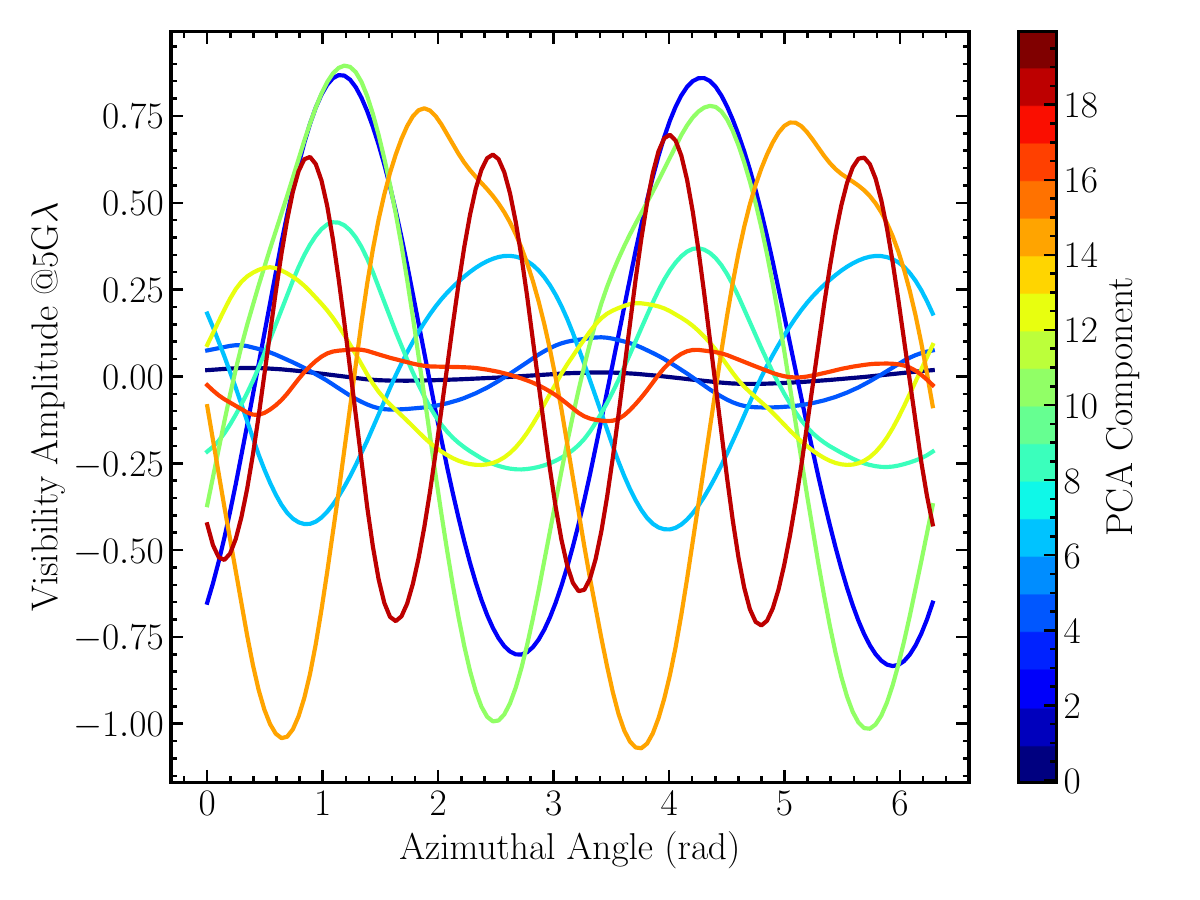}
\includegraphics[width=0.45\linewidth]{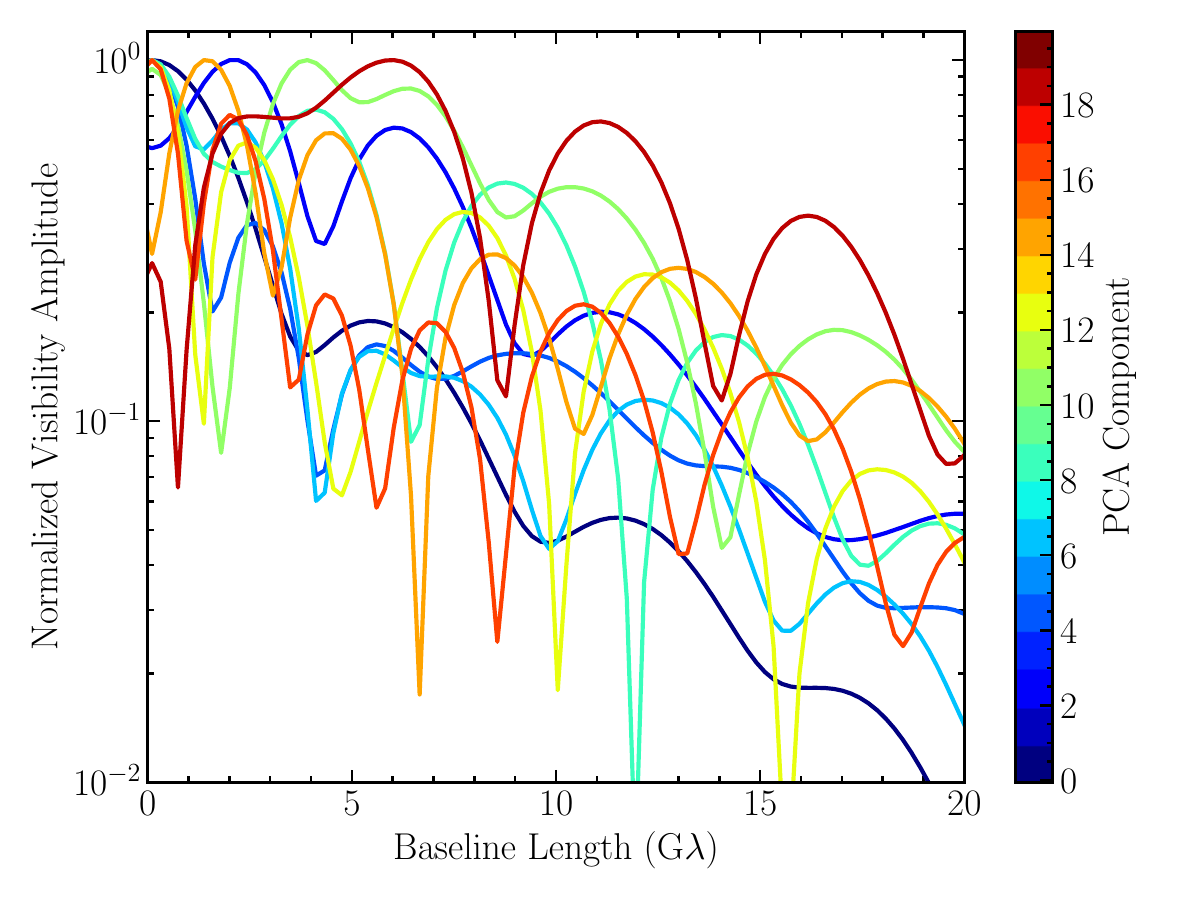}
    \caption{The {\em (Left)\/} azimuthal and {\em (Right)\/} radial dependence of the Fourier maps of the low-order eigenimages trained on simulated black-hole images. In the left panel, we evaluated the eigenimages at a baseline length of 5~G$\lambda$, which corresponds to the maximum of the secondary peak in Fourier spectrum of the M87 black hole. In the right panel, we evaluate the eigenimages along a cross section perpendicular to the spin axis of the black hole. For brevity, we only show the even eigenimages and, in the left panel, we have also subtracted the mean visibility amplitude of each eigenimage. The dense coverage in the locations of nodes and minima in these eigenimages provide an empirical justfication of their generality and near completeness.}
    \label{fig:eigen}
\end{figure*}

There is, of course, an infinite set of choices for the set of eigenimages (or, equivalently, for the set of eigenmaps in the Fourier domain) that one could use to reconstruct the sparsely populated Fourier plane and, hence, the corresponding image. In principle, we can decompose the Fourier plane into a linear combination of products of azimuthal $e^{i m \phi}$ modes and a complete set of radial functions and use the data to infer the coefficients of this linear decomposition. Examples of such decompositions include the Legendre and the Zernike polynomials or more complex functions such as wavelets~\citep{Aghabiglou2024}. Even though these are different sets of eigenfunctions, they are complete and can be used interchangeably in describing any two-dimensional function. The key difference between them is the rate of convergence of the linear decomposition for a given type of image. If the convergence is fast and only a small number of eigenimages are necessary to describe the observed image structures, then the reconstruction can proceed without the need of imposing regularizers. If, instead, the convergence is slow, then {\em ad hoc} regularizers will need to be imposed (see, e.g, image reconstruction with wavelet decomposition that requires regularizers;~\citealt{Aghabiglou2024}). Therefore, the choice that optimizes the compactness of the decomposition depends on the particular properties of the images under study.

We are interested in reconstructing images of black holes with two important characteristics that can be directly inferred from the data (see Fig.~\ref{fig:eht}). First, the images have a finite extent, as can be seen by the observed visibility amplitudes that drop with a characteristic baseline length of 2~G$\lambda$. As a consequence of this finite size, the optimal radial eigenfunctions in the Fourier domain also need to decrease with baseline length. This makes the most common orthonormal polynomials (e.g., Legendre, Laguerre, etc) suboptimal for fast radial convergence of the Fourier maps.

Second, the Fourier maps of the images have a range of deep minima at baseline lengths that, for the case of the M87 black hole, are multiples of $\sim 3.5$~G$\lambda$, as it can again be inferred directly from the data (Fig.~\ref{fig:eht}). Such minima are smoking guns of images with sharp edges, such as disks and rings. This is consistent with the theoretical expectation that black-hole images at millimeter wavelengths have sharp edges at the boundaries of black-hole shadows~\citep{Psaltis2015}. It is straightforward to show that, if we were to reconstruct the Fourier maps with linear combinations of some of the common orthonormal polynomials, the resulting images would not have sharp edges.

In order to overcome these difficulties, we opted instead to use an algorithm that ``learns'' the optimal polynomials from a training set of images. This is the context within which we consider \texttt{PRIMO} a machine learning algorithm. As discussed in the introduction, among different sparse dictionary learning approaches, we opted to use a Principal Component decomposition because of its linearity and the orthogonality of the eigenimages. Decomposing images to PCA components has been used extensively in computer vision for decades~\citep[see, e.g.,][]{Turk1991}. In \citet{Medeiros2018}, we demonstrated that a PCA decomposition of black-hole images obtained from high-resolution GRMHD simulations provides a basis set of eigenimages with very fast convergence: only twenty eigenimages are required to account for more than 99\% of the variance observed in {\em tens of thousands} of black-hole snapshots.

However, an important question remains. Are these eigenimages sufficient to describe images that are not part of the training set and would the convergence of that description be also fast? This question was explored partially in \citet{Medeiros2023a}, where black-hole images from simulations obtained with one set of assumptions regarding the underlying physics were reconstructed successfully using training images from simulations with a different set of physical assumptions. Furthermore, \citet{Hallur2022} showed that images of black holes can be reconstructed (albeit requiring more components) {\em even when the algorithm is trained on images of red-noise structure within a finite image size.}

These experiments suggest that, as long as an extensive training set of images has been used, the details of the simulations used to generate them are of less importance. Figure~\ref{fig:eigen} provides additional empirical justification to this claim. The left panel shows the angular dependence in the Fourier maps of the low-order eigenimages, evaluated at a given baseline length. We used here a baseline length of 5~G$\lambda$, which corresponds to the maximum of the secondary peak in the Fourier spectrum of the M87 black hole (see Fig.~\ref{fig:eht}). It is clear that the azimuthal functional forms of the eigenimages are very similar to $e^{\i m\phi}$ polynomials with different azimuthal ``wavelengths'' (or number of nodes), which themselves form a complete basis set.

The right panel of Figure~\ref{fig:eigen} shows the radial dependence of the same eigenimages. All eigenimages show the two characteristics of black-hole images that we could see directly in the M87 data: visibility amplitudes that decrease with increasing baseline length and deep minima. However, the locations of these minima, which quantify the radial ``wavelengths'' of ringing of these eigenfunctions, provide a dense coverage in the range of baseline lengths that are of interest. 

It is this broad range of possibilities encoded in the eigenimages learned by the algorithm together with the presence of high-quality data that are separated by distances in Fourier space that are smaller than the characteristic scale of variation that allow for \texttt{PRIMO} to recreate a robust image that depends only marginally on the training set used.

\section{Discussion}

In this paper, we used a variety of analytical and numerical arguments to explore the foundations of algorithms such as \texttt{PRIMO} that reconstruct images of cosmic sources based on sparse interferometric data. A fundamental concept underpinning this exploration is the recognition that the Fourier components of compact images are highly correlated within regions of size equal to the baseline length that resolves the image. As a result, if the size of the gaps in the interferometric data are at most comparable to the resolving baseline length, an algorithm with sufficient flexibility can reconstruct the underlying image without the necessity to populate the data gaps with arbitrary information.

Even though our work is focused on understanding the dictionary learning algorithm \texttt{PRIMO}, it also provides insights into other machine-learning imaging approaches that utilize convolutional neural networks~\cite[CNNs;][]{Schmidt2022,Connor2022} or generative adversarial networks~\cite[GANs;][]{Geyer2023}. The goal of these studies has been to use deep learning approaches in order to achieve super-resolution, i.e., resolve images at finer scales than the nominal resolution of the interferometer~\citep[see also, e.g., ][for similar aims within regularized maximum likelihood methods]{Honma2014}. Our work demonstrates that the largest baseline length for which the data can provide sufficient information to any algorithm for the structure of the underlying image is equal to the sum of two terms: {\em (i)} the largest baseline length $b_{\rm max}$ at which the interferometer provides good coverage and {\em (ii)} the baseline length $b_0$ that resolves the image that is being observed. If we denote the nominal resolution of the array by $\theta_{\rm res}\sim 1/b_{\rm max}$ and the angular size of the image by $\theta_{\rm im}\sim 1/b_0$, then the finest structures that can be resolved by any algorithm that does rely heavily on prior knowledge of or restrictive assumptions on the underlying image will have an angular size equal to
\begin{equation}
\theta_{\rm min} \gtrsim \left(\frac{1}{\theta_{\rm res}}+\frac{1}{\theta_{\rm im}}\right)^{-1}\;.
\end{equation}
Structures with angular sizes smaller than this limit cannot be inferred from the data, whether one uses machine learning algorithms or not, and will necessarily require generating information that is not constrained by observations.

It is important to emphasize here that \texttt{PRIMO} only addresses the issue of image reconstruction from data obtained with a sparse interferometer. It does not attempt to improve the calibration of the data in any way other than introducing an unknown time-dependent gain for each station in the array, which is inferred as part of the fitting procedure. This is different than machine-learning algorithms such as \texttt{AIRI}~\citep{Dabbech2022}, which have been developed in order to refine the calibration of data obtained with radio interferometers. 

For the specific application of black-hole imaging with sparse interferometry, we provided quantitative arguments using the 2017 EHT array with 5 geographical locations. However, the methods we employ are applicable and informative even if the number of geographical locations increases. Indeed, the EHT array is continuing to evolve with the addition of sites such as the Greenland Telescope (GLT) in 2018, and the Kitt Peak Telescope in Arizona and NOEMA in the French Alps in 2022. Even though the presence of these additional sites is a step towards reducing partially the sparseness of the array, the maximum baseline lengths, which are limited by the size of the Earth, are not affected. Moreover, even in the 2022 array there remain data gaps of size comparable to the baseline lengths that resolve the images of the black holes in the center of the M87 and the Milky Way galaxies. We, therefore, expect \texttt{PRIMO} to continue to provide efficient and high-fidelity reconstructions of these black-hole images using EHT data obtained in subsequent years.

\begin{acknowledgments}

We thank the members of the Xtreme astrophysics group at Georgia Tech and at the University of Arizona for useful discussions and comments. This work was supported in part by NSF PIRE grant OISE-1743747. L.\;M.\ gratefully acknowledges support from a NASA Hubble Fellowship Program, Einstein Fellowship under award number HST-HF2-51539.001-A.

\end{acknowledgments}

\mbox{}

\bibliographystyle{aasjournal}

\bibliography{primo}{}

\end{document}